\documentclass[prd,aps,twocolumn,nofootinbib,preprintnumbers]{revtex4}
\usepackage{graphicx}

\begin{document}

\newcommand{\gsim}{ \mathop{}_{\textstyle \sim}^{\textstyle >} }
\newcommand{\lsim}{ \mathop{}_{\textstyle \sim}^{\textstyle <} }
\newcommand{\vev}[1]{ \left\langle {#1} \right\rangle }

\newcommand{\bear}{\begin{array}}  \newcommand{\eear}{\end{array}}
\newcommand{\bea}{\begin{eqnarray}}  \newcommand{\eea}{\end{eqnarray}}
\newcommand{\beq}{\begin{equation}}  \newcommand{\eeq}{\end{equation}}
\newcommand{\bef}{\begin{figure}}  \newcommand{\eef}{\end{figure}}
\newcommand{\bec}{\begin{center}}  \newcommand{\eec}{\end{center}}
\newcommand{\non}{\nonumber}  \newcommand{\eqn}[1]{\beq {#1}\eeq}
\newcommand{\la}{\left\langle} \newcommand{\ra}{\right\rangle}

%
%%%%%%%%%%%%%%%%%%%%%%%%%%%%%%%%%%%%%%%%%%%%%%%%%%%%%%%%%%%%%%%%%%%%

\renewcommand{\thefootnote}{\alph{footnote}}

\renewcommand{\thefootnote}{\fnsymbol{footnote}}
\preprint{DESY 06-014}
\title{Moduli-Induced Gravitino Problem}
\renewcommand{\thefootnote}{\alph{footnote}}

\author{Motoi Endo$^{1,2}$, Koichi Hamaguchi$^1$ and Fuminobu Takahashi$^{1,2}$}

\affiliation{
${}^1$ Deutsches Elektronen Synchrotron DESY, Notkestrasse 85,
  22607 Hamburg, Germany\\
${}^2$ Institute for Cosmic Ray Research,
  University of Tokyo, Chiba 277-8582, Japan
  }

\begin{abstract}
\noindent
We investigate the cosmological moduli problem by studying a modulus
decay in detail and find that the branching ratio of the gravitino production
 is generically of $O(0.01-1)$, which causes another
cosmological disaster.
Consequently, the cosmological moduli problem cannot be solved simply
by making the modulus mass heavier than $100\,{\rm TeV}$. We also
illustrate our results by explicitly calculating the branching ratio 
into the gravitinos in the mixed modulus--anomaly/KKLT- and racetrack-type
models.
\end{abstract}

\maketitle

%%%%%%%%%%%%%%%%%%%%%%%%%%%%%%%%%%%%%%%%%%%%%%%%%%%%%%%%%%%%%%%%%%%%
%\section{Introduction}

The cosmological moduli problem~\cite{ModuliProblem} is one of the
most challenging puzzles in particle physics and cosmology. In this
letter, we show that the problem is even more difficult than usually
thought.

In supergravity/superstring theories, generically there exist moduli
fields which have flat potentials and obtain masses from
supersymmetry (SUSY)  breaking and non-perturbative effects.  
During an inflationary period, a modulus field $X$ is likely to 
develop a large expectation value. After the end of the inflation, 
it starts a coherent oscillation and soon dominates the energy density 
of the universe. Due to the interaction suppressed by the Planck scale 
$M_P=2.4\times 10^{18}$ GeV, the decay rate of the modulus $X$ is 
extremely small:
\begin{eqnarray}
  \Gamma_X &=&  {c\over 4\pi} {m_X^3 \over M_P^2}\;,
  \label{eq:GammaX}
\end{eqnarray}
which leads to an onset of a radiation--dominated universe with a very
low temperature:
\begin{eqnarray}
  T_X &=& (\pi^2 g_*/90)^{-1/4}\sqrt{M_P \Gamma_X}
  \nonumber\\
  &\simeq& 5.5\times 10^{-3}~\mathrm{MeV}
  \cdot
  c^\frac{1}{2}\left({m_X \over 1~\mathrm{TeV}}\right)^{3/2}\;.
  \label{eq:Tx}
\end{eqnarray}
Here, $c$ is an order one coefficient and $g_*$ is the effective
number of massless degrees of freedom. This is cosmologically
unacceptable because a successful big--bang nucleosynthesis (BBN)
requires that the (last) radiation--dominated universe starts with
temperature higher than $\sim 5$ MeV~\cite{Kawasaki:1999na}.

As is clear from Eq.~(\ref{eq:Tx}), a simple solution would be to
assume that the modulus $X$ is ultra heavy~\footnote{
See Refs.~\cite{other-sol,Yokoyama:2006wt} for other solutions.
}:
\begin{eqnarray}
  m_X \;\gsim\; 100~\mathrm{TeV}
  \quad
  \to
  \quad
  T_X \;\gsim\; \mathcal{O}(\mathrm{MeV})\;.
  \label{eq:mX100TeV}
\end{eqnarray}
Actually, there have been proposed scenarios with such a
large modulus mass (cf.~\cite{AM,KKLT,Mirage,racetrack,Buchmuller:2004xr}). 

However, there exists yet another serious cosmological obstacle even
for heavy moduli scenarios.  The new problem is caused by the gravitino
which is produced by the modulus decay.  Indeed, as we will show, the
branching ratio of the modulus decay into the gravitino is generically
quite large
\begin{eqnarray}
  B_{3/2}
  \;\equiv\;
  \mathrm{Br} (X \to \mathrm{gravitino}) 
  \;=\; \mathcal{O}(0.01-1)\;,
  \label{eq:BrX}
\end{eqnarray}
which causes serious problems after the modulus decay. 
We call this problem the {\it moduli--induced gravitino problem}. 

The gravitino production via modulus decay
and its cosmological implications have been previously discussed in 
Refs.~\cite{Hashimoto:1998mu,KYY}, taking $\mathrm{Br}
(X\to \mathrm{gravitino})\ll 1$.  The main purpose of this letter is
to show that Eq.~(\ref{eq:BrX}) holds in a generic setup, and to
emphasize how disastrous its consequences are. We also exemplify
explicit results in the mixed modulus--anomaly/KKLT mediation~\cite{KKLT,Mirage}
and in the racetrack~\cite{racetrack} setups.

Let us first estimate the branching ratio of a modulus decay into
gravitino(s). We consider a heavy modulus scenario, $m_X\gsim
100~\mathrm{TeV}$ [cf. Eq.~(\ref{eq:mX100TeV})]. On the other hand,
the gravitino is likely to be (much) lighter than 100~TeV, because too
large gravitino mass requires a fine-tuning in the Higgs sector due to
the anomaly-mediated effects. Thus, we assume $m_X \gg m_{3/2}$
hereafter.  After choosing the unitary gauge in the Einstein frame, 
where the goldstino component is completely absorbed into the gravitino 
under the super--Higgs mechanism~\cite{Cremmer:1978hn}, the relevant 
gravitino-modulus couplings are~\cite{Hashimoto:1998mu,KYY,WessBagger}
\bea
  e^{-1}\mathcal{L} &=&
  - \frac{1}{8} \epsilon^{\mu\nu\rho\sigma} 
  \left( G_X \partial_\rho X 
    - G_{\bar X} \partial_\rho \bar X \right)
  \bar \psi_\mu \gamma_\nu \psi_\sigma\non\\
  &&
  - \frac{1}{8} e^{G/2} \left( G_X X + G_{\bar X} \bar X \right)
  \bar\psi_\mu \left[\gamma^\mu,\gamma^\nu\right] \psi_\nu, 
\eea
where $\psi_\mu$ is the gravitino field. 
Here and hereafter, we set $M_P=1$ unless explicitly written.  
The function $G = K + \ln |W|^2$ is the total K\"ahler potential, 
and $G_i$ denotes a derivative of $G$ with respect to the field $i$.  
The real and imaginary components of $X = (X_R + iX_I)/\sqrt{2}$ have 
the decay rate~\footnote{
  The branching ratio of the single gravitino production is suppressed 
  due to the phase space; ${\rm Br}(X \rightarrow \tilde X \psi_{3/2}) 
  \lesssim (m_{3/2}/m_X)^2$.
},
\begin{eqnarray}
  \Gamma(X_{R, I} \rightarrow 2\psi_{3/2}) \;\simeq\;
  \frac{1}{288\pi} \frac{|G_X|^2}{g_{X\bar X}} \frac{m_X^5}{m_{3/2}^2}, 
  \label{eq:gamma_gravitino_GX}
\end{eqnarray}
in the limit of $m_X \gg m_{3/2}$ after the canonical normalization, 
$\hat X = \sqrt{g_{X\bar X}} X$, where $g_{ij^*}=K_{ij^*}$ is the K\"ahler metric. 
We emphasize that the gravitino mass in the denominator arises from 
the longitudinal component of the gravitino, which corresponds to the goldstino, 
$\psi_\mu \sim (p_\mu/m_{3/2}) \chi_{GS}$, in the massless limit.  
This factor magnifies the decay rate by a factor of $\sim (m_X/m_{3/2})^4$, 
compared to the previous 
results~\cite{Hashimoto:1998mu,KYY,Moroi:1999zb}.

The auxiliary field of the modulus, $G_X$, represents its fractional
contribution to the total amount of the SUSY breaking. It
may be small enough to
suppress the gravitino-production rate. However, we can see that in
the framework of the 4D $N=1$ supergravity, $G_X$ is generically
bounded below, $G_X\gsim m_{3/2}/m_X \equiv R^{-1}$.  The scalar
potential is given by $V = e^G(G^i G_i - 3)$ and the (almost) vanishing
vacuum energy demands that at least one of the auxiliary fields,
$G_i$, should take $G^i G_i \sim O(1)$, where $G^i = g^{ij^*}
G_{j^*}$.  If the modulus field plays the role, it dominantly decays
into the gravitino [cf. Eqs.(\ref{eq:GammaX}) and
(\ref{eq:gamma_gravitino_GX})] and hence clearly $B_{3/2}\simeq 1$.
Instead, by introducing a hidden sector field $Z$ with $G_Z \sim
O(1)$, the modulus-auxiliary field $G_X$ can be small~\footnote{
The potential may be up-lifted by adding the $D$-term, $V^{(D)}$.
One can show that $|G_X| \gtrsim R^{-1}$ also in this case, 
as far as $V^{(D)}_X/V^{(D)} \sim O(1)$, which corresponds to
the second term in Eq.~(\ref{eq:st-X}).
}. In the
following discussion, we assume that the modulus field does not decay
into the hidden-sector field for simplicity. In order to see how small
$G_X$ can be, we investigate the condition to minimize the potential, 
$V_X = 0$, leading to
\begin{eqnarray}
\label{eq:st-X}
  G^X \nabla_X G_X + G^Z \nabla_X G_Z + G_X \;=\; 0,
\end{eqnarray}
where $\nabla_i G_j = G_{ij} - \Gamma^k_{ij} G_k$ with the connection,
$\Gamma^k_{ij}=g^{k\ell^*}g_{i \ell^{*} j}$.  We assume that the K\"ahler potential 
includes no enhancement factor, and especially $g_{X\bar X},\,g_{Z\bar Z} \sim 1$ 
and $K_X \sim 1$. The first term in Eq.~(\ref{eq:st-X}) then becomes 
$\sim G^X R$, because $m_X/m_{3/2} \sim |\nabla_X G_X| + 
\mathcal{O}(G_i)$.  Barring cancellation, the second term is
given by ${\rm max}\left[W_Z/W,\,W_{XZ}/W, K_{XZ}, \Gamma^Z_{XZ}\right]$, 
which is $O(1)$ unless the hidden sector takes a special form.  
Thus, we arrive at $|G_X| \sim R^{-1}$ for $|g^{X\bar Z}| < R^{-1}$ and 
$|G_X| \gtrsim R^{-1}$ for $|g^{X\bar Z}| \gsim R^{-1}$.
As a result, Eq.~(\ref{eq:gamma_gravitino_GX}) becomes
\begin{eqnarray}
  \Gamma(X_{R,I} \rightarrow 2\psi_{3/2}) \;\simeq\;
  \frac{|\kappa|^2}{288\pi} \frac{1}{g_{X\bar X}} m_X^3,
  \label{eq:gamma_gravitino}
\end{eqnarray}
where we define $G_X = \kappa/R = \kappa (m_{3/2}/m_X)$ with $|\kappa| \gsim 1$. 
Note that the above discussion is valid for any value of $m_X$,
as long as $R \gg 1$ is satisfied.

The modulus field also decays into radiation, that is, the standard
model (SM) particles and their superpartners. 
The relevant interactions of $X$ then 
stem from the dilatonic coupling with the gauge sector, $\int d^2\theta\,X WW$, 
leading to
\begin{eqnarray}
e^{-1}\mathcal{L} &=& 
  -\frac{1}{4\sqrt{g_{X\bar X}}}
  \bigg[ \frac{\hat{X}_R}{ \langle X_R \rangle} F_{\mu\nu}^{(a)} F^{(a) \mu\nu} +
  \frac{ {\hat X}_I}{ \langle X_R \rangle} F_{\mu\nu}^{(a)} \widetilde{F}^{(a) \mu\nu} 
   \non\\
  && 
\left.-\frac{\sqrt{2}}{\la X_R \ra} e^{G/2}
 \left(G^{\bar{X}}_{\phantom{a}X}  \hat{X} \,\bar{\lambda}^{(a)} {\mathcal P}_R \lambda^{(a)}
+ {\rm h.c.}
\right)\right]
\end{eqnarray}
after the canonical normalization, where $F_{\mu\nu}$ and $\lambda$
are the field strength of the gauge boson and its superpartner with 
a generator index $a$ of the corresponding gauge symmetry.  The chiral 
projection operators are defined as ${\mathcal P}_{R,L} = (1\pm\gamma_5)/2$. 
We notice $e^{G/2} |G_{\phantom{a}X}^{\bar X}|$ is approximately given by 
the modulus mass for $R \gg 1$.  Then the decay rate is
\begin{eqnarray}
\label{eq:decay-rate-gauge}
  \Gamma(X_{R,I} \rightarrow radiation) \;\simeq\;
  \frac{3}{8\pi} \frac{1}{g_{X\bar X}} \frac{m_X^3}{\langle X_R \rangle^2},
  \label{eq:gamma_radiation}
\end{eqnarray}
for $SU(3)_c \times SU(2)_L \times U(1)_Y$, corresponding to $c = O(1)$. 
We notice that a half of 
the decay rate comes from the channel of the gaugino production, in contrast 
to the results of Refs.~\cite{Moroi:1999zb,KYY}.
From Eqs.~(\ref{eq:gamma_gravitino}) and (\ref{eq:gamma_radiation}),
we obtain the branching ratio of the gravitino production as
\begin{eqnarray}
  {\rm Br}(X_{R,I} \rightarrow 2\psi_{3/2}) \;\simeq\; 
  \frac{|\kappa|^2\langle X_R \rangle^2/108}
  {1 + |\kappa|^2\langle X_R \rangle^2/108}.
\end{eqnarray}
It is important to note that the production rate of the gravitino channel 
is one of the dominant processes in the  modulus decay. Actually, it becomes 
an order of $0.01-1$. 

The other decay processes are suppressed,
except for a possible decay into the Higgs(-ino)~\cite{Moroi:1999zb}. However, 
this decay channel is model-dependent, and does not change the above result
$c = O(1)$.

%\subsection{Cosmological Bound}

Let us show how such large branching ratio into the gravitinos
jeopardizes the success of the standard big bang cosmology.  
We consider the constraints from (i) the speed-up effect, (ii) BBN, and
(iii) the lightest SUSY particle (LSP) abundance.
First
let us consider the so-called speed-up effect which modifies mostly
the $^4$He abundance.  This sets a bound on the abundance of the
gravitinos with $m_{3/2} \lesssim 20$ TeV (cf.~\cite{Kohri:2005wn}), since they
decay after the neutron-proton transformation decouples.  The
observational data put an upper bound on the ratio of the energy densities
of the gravitinos to the standard model particles at the BBN
epoch~\cite{Cyburt:2004yc} as 
 \beq
 \label{eq:rho-ratio-obs}
f_{3/2} \equiv \left.\frac{ \rho_{3/2}}{\rho_{SM}}\right|_{\rm BBN} \lesssim
0.2,~~(95\% {\rm C.L.}).
\eeq
In our heavy moduli scenario, however, the ratio is bounded below:
$f_{3/2}  \geq B_{3/2}/(1-B_{3/2})$, 
where the equality holds if the gravitinos are still relativistic at
the BBN epoch and most of the superpartner of the SM particles
directly produced from the modulus decay soon annihilate into the SM
particles. Thus we obtain
\beq
\label{eq:speed-up}
B_{3/2} < 0.2,
\eeq
irrespective of whether the gravitino is stable or unstable.
For $B_{3/2} > 0.2$, the gravitinos from the modulus decay always
upset the standard BBN, as long as $m_{3/2} \lesssim 20{\rm\,TeV}$.

Now we discuss the cases of the stable and unstable gravitinos separately. 
First we take up the unstable gravitinos, which is the case if the gravitino
mass is heavier than the LSP mass, $m_{3/2} >
m_{\rm LSP}$.
The gravitino-to-entropy ratio is given by
\bea
Y_{3/2} &\equiv& \frac{n_{3/2}}{s} \simeq 2 B_{3/2} \frac{3 T_X}{4 m_X},\non\\
&\simeq& 2.6 \times 10^{-7}\,c^\frac{1}{2} B_{3/2} 
\left(\frac{m_X}{10^3~{\rm TeV}}\right)^\frac{1}{2}.
\eea
The BBN severely constrains $Y_{3/2}$~\cite{Kawasaki:2004yh, Kohri:2005wn}.
Even if we adopt the recent analysis on $^4$He abundance which has taken account
of possible large systematic error~\cite{Olive:2004kq}, $Y_{3/2}$
must be smaller than $2 \times 10^{-12}$ at $95\%$ C.L. for
$m_{3/2} \simeq 30$ TeV~\cite{Kohri:2005wn}, and the bound becomes much severer 
for smaller $m_{3/2}$.  Therefore the branching ratio into the
gravitinos must be extremely small: 
\beq
\label{eq:bbn-limit}
B_{3/2} <10^{-5}   \epsilon\,c^{-\frac{1}{2}} \left(\frac{10^3~{\rm TeV}}{m_X}
\right)^\frac{1}{2},
\eeq
for $m_{\rm LSP}<m_{3/2} \leq 30{\rm\,TeV}$. Here $\epsilon \leq 1$
parameterizes the BBN bound: $\epsilon =1$ for $m_{3/2} \simeq
30{\rm\,TeV}$, and $10^{-5} \lesssim \epsilon \ll 1$ for $m_{3/2} <
30{\rm\,TeV}$.  In addition, the abundance of the LSPs from the
gravitino decay is (cf.~\cite{Fujii:2001xp})
\beq
\label{eq:YLSP}
\left.Y_{{\rm LSP}}\right|_{\psi_{3/2}}\simeq  {\rm min}\left[ Y_{3/2}, 
  \sqrt{\frac{45}{8\pi^2g_*}} \frac{1}{ M_P  T_{3/2}\la \sigma_{ann} v \ra }
\right],
\eeq
where $\la \sigma_{ann} v \ra$ is the thermally averaged annihilation 
cross section
of the LSP, and $T_{3/2}$ is the decay temperature of the gravitino.
Since the LSP abundance must be smaller than the dark matter abundance,
we have another constraint on $B_{3/2}$: 
\beq
\label{eq:LSP-from-g}
B_{3/2} < 1.8 \times10^{-5} c^{-\frac{1}{2}}  
\left(\frac{100~{\rm GeV}}{m_{\rm LSP}}\right) 
\left(\frac{\Omega_{\rm DM}h^2}{0.13}\right) 
\left(\frac{10^3~{\rm TeV}}{m_X}\right)^\frac{1}{2}
\eeq
for $m_{3/2}>m_{\rm LSP}$. Here
$\Omega_{\rm DM}$ is the density parameter of the dark matter, $h$ is
the present Hubble parameter in units of $100$km/sec/Mpc, and we have
assumed $\la \sigma_{ann} v \ra < 10^{-6}$ GeV$^{-2}$ and $m_{3/2} < 100$ TeV.
We can see that (\ref{eq:speed-up}), (\ref{eq:bbn-limit}) and 
(\ref{eq:LSP-from-g})
rule out the unstable gravitinos, unless $B_{3/2}$ is 
extraordinarily small in spite of our result $B_{3/2} = O(0.01-1)$.

\begin{figure}[t]
\begin{center}
\includegraphics[width=7cm]{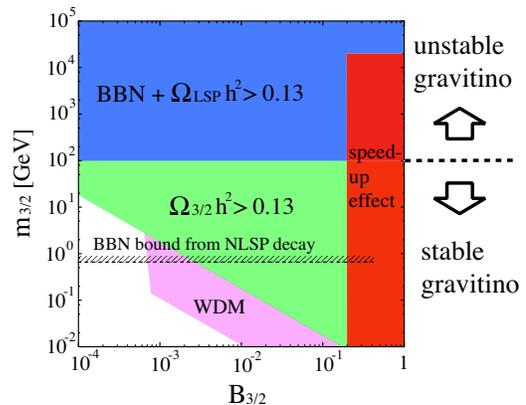}
\caption{The cosmological bounds on $m_{3/2}$ and $B_{3/2}$. Shaded regions are
  excluded by cosmological arguments. See the text for details. 
  The horizontal dashed line denotes the BBN bound from the stau NLSP 
  decay into gravitinos for $m_{\rm NLSP}=100$ GeV. 
  We have chosen $m_X = 10^3$ TeV
  and $c=1$ as representative values. The bounds become severer for larger $m_X$.}
\label{fig:mg-br}
\end{center}
\end{figure}

Next we consider the stable gravitinos, which is the case if the gravitino
is the LSP. A constraint then comes  from the requirement that the
gravitino abundance should not exceed the dark matter abundance,
and we only have to replace $m_{\rm LSP}$ with $m_{3/2}$ in  (\ref{eq:LSP-from-g}):
\beq
\label{eq:g-dm}
B_{3/2} < 1.8 \times10^{-2} c^{-\frac{1}{2}}   
\left(\frac{100~{\rm MeV}}{m_{3/2}}\right) 
\left(\frac{\Omega_{DM}h^2}{0.13}\right) 
\left(\frac{10^3~{\rm TeV}}{m_X}\right)^\frac{1}{2}
\eeq
for the stable gravitino. Furthermore, the bound on the gravitino 
abundance is severer by an order of magnitude for 
$m_{3/2} \lesssim 100~{\rm MeV} (m_X/10^3~{\rm TeV})^{-1/2}$, 
due to the present large free-streaming
velocity~\cite{Jedamzik:2005sx}. See Fig.~\ref{fig:mg-br}. 
From (\ref{eq:speed-up}) and (\ref{eq:g-dm}), we conclude that large $B_{3/2}$
encounters trouble even for the stable gravitinos. 

Lastly we comment on the lightest superpartner of the SM particles, 
denoted by $\chi$, produced from the modulus decay through the
gauginos. From the discussion above, the number of $\chi$
generated from the decay of unit quantum of the modulus field is order unity.
If $\chi$ is the LSP and stable, it must be electrically neutral.
To satisfy $\Omega_{\rm LSP} h^2 <0.13$,
the pair annihilation 
cross section must be large~\cite{Fujii:2001xp,Moroi:1999zb}, 
which in turn constrains the 
mass spectrum of the SUSY particles. 
In the case of the gravitino LSP, there is a strict BBN bound on 
the abundance and lifetime of the next-to-LSP (NLSP) $\chi$.
For a stau NLSP $\tilde{\tau}$, we find upper bounds on the
gravitino mass, $m_{3/2} \lesssim (0.3-1)$ GeV for $m_X = 10^3$ TeV and
$100\ {\rm GeV}\lesssim m_{\tilde{\tau}} \lesssim 1$ TeV~\footnote{
  We have used $Y_{\tilde \tau} \simeq (45/8\pi^2g_*)^{1/2} (M_P T_{X}\la 
  \sigma_{ann} v \ra)^{-1}$~\cite{Fujii:2001xp} with  
  $\la \sigma_{ann} v\ra \lesssim 10^{-3}m_{\tilde{\tau}}^{-2}$~\cite{Asaka:2000zh}, 
  BBN bounds from \cite{Kohri:2005wn} with 
  $\tilde{\tau}$'s hadronic branching in \cite{Feng:2004zu},
  and $\tau_{\tilde{\tau}} \simeq 6{\rm \,sec}\,(m_{\tilde{\tau}}/100~{\rm GeV})^{-5}
  (m_{3/2}/10{\rm\,MeV})^2$. The bound from $\Omega_{3/2}$ (from $\tilde{\tau}$) 
  $<\Omega_{\rm DM}$ is slightly weaker than the BBN bound.
}.   For a neutralino NLSP, the bound becomes severer.

%We summarize the bounds  (\ref{eq:speed-up}), (\ref{eq:bbn-limit}),
%(\ref{eq:LSP-from-g}) and (\ref{eq:g-dm})  in Fig.~\ref{fig:mg-br},
%from which we can see how serious the cosmological moduli problem
%becomes as a result of our finding that $B_{3/2}$ should be $O(0.01-1)$.  
%Note that the bounds on $B_{3/2}$ become severer for heavier
%modulus mass and larger $c$, although the dependence is weak.

We summarize the bounds considered above in Fig.~\ref{fig:mg-br},
from which we can see how serious the cosmological moduli problem
becomes as a result of our finding that $B_{3/2}$ should be $O(0.01-1)$.  
Note that the bounds on $B_{3/2}$ [cf.  (\ref{eq:bbn-limit}),
(\ref{eq:LSP-from-g}) and (\ref{eq:g-dm})] become severer for heavier
modulus mass and larger $c$, although the dependence is weak.

%%%%
%%%%

%\subsection{Examples}

Let us explicitly calculate the 
branching fraction of the gravitino in some models.
The relevant terms in the Lagrangian of the mixed modulus--anomaly/KKLT 
mediation model~\cite{KKLT,Mirage} are given by
\begin{eqnarray}
  \label{eq:KKLT_Lagrangian}
  && f \;=\; -3(X+\bar{X})^{n/3} + C_Z(X+\bar{X})^l |Z|^2, \\
  && W \,=\, W_0 + a e^{-bX} + W_{\rm Hidden}(Z),
\end{eqnarray}
where the K\"ahler potential is defined as $f = -3e^{-K/3}$. 
Here coefficients $b$ and $C_Z$ are  real with $bX_R \gg 1$,  while $a$ and
$W_0$  are complex. Then the modulus mass is
\begin{eqnarray}
  m_X^2 \;\simeq\; 2 (b X_R)^2 m_{3/2}^2.
\end{eqnarray}
The gravitino mass $m_{3/2}$ is much larger than 
the soft mass scale~\cite{Mirage}.
On the other hand, $G_X$ is
\begin{eqnarray}
  G_X \;\simeq\; -\frac{2n + 3l}{2 X_R} \frac{1}{bX_R},
\end{eqnarray}
where we assumed $\langle Z \rangle \ll 1$ and the vanishing cosmological constant. 
Thus the branching ratio becomes
\begin{eqnarray}
  {\rm Br}(X_I \rightarrow 2\psi_{3/2}) \;\simeq\; 
  \frac{(2n + 3l)^2/216}
  {1 + (2n + 3l)^2/216}.
  \label{KKLT_Br}
\end{eqnarray}
In the original KKLT model~\cite{KKLT} with the lift-up potential due to 
$\overline{D3}$ brane, which is realized by choosing $n = 3$ and $l = 0$,
the branching ratio is $B_{3/2} \simeq 0.14$.

In the racetrack-type models~\cite{racetrack}, the K\"ahler potential is 
the same as Eq.~(\ref{eq:KKLT_Lagrangian}) but the superpotential is
\begin{eqnarray}
  W \,=\, a_1 e^{-b_1X} + a_2 e^{-b_2X} + W_{\rm Hidden}(Z),
\end{eqnarray}
where $a_i$'s are complex and $b_i$'s are real numbers. Here $b_i X_R\gg 1$. 
The modulus mass and $G_X$ are 
\begin{eqnarray}
  && m_X^2 \;\simeq\; \frac{4}{n^2} (b_1 X_R)^2 (b_2 X_R)^2 m_{3/2}^2, \\
  && G_X   \;\simeq\; 
  -\frac{n(2n + 3l)}{2\sqrt{2} X_R} \frac{1}{(b_1X_R)(b_2X_R)}.
\end{eqnarray}
Therefore the branching ratio becomes the same as 
Eq.~(\ref{KKLT_Br}). For $n=1$ and $l=0$, one obtains $B_{3/2} \simeq 0.018$.

%%%%%%%%%%%% 
% \section{summary}
%%%%%%%%%%%% 
In summary, we have shown that the branching ratio of the modulus decay
into gravitinos is $O(0.01-1)$, and hence heavy 
modulus scenario is plagued with the moduli-induced gravitino problem. 
Here let us comment on possible ways out. 
Unless either $B_{3/2}$ or the gravitino mass
is extremely small,
there is no way to circumvent 
the problem other than introducing something that dilutes
the modulus field and the subsequently produced gravitinos and (N)LSPs.
One of the candidates is the thermal inflation~\cite{thermal-inf}. 
Another is Q-balls~\cite{Coleman}
in the Affleck-Dine mechanism~\cite{AD}. In the latter case, 
the Q-balls can not only 
dilute the unwanted gravitinos but also generate the baryon
asymmetry successfully. Detailed discussion is beyond the scope of 
this letter and will be presented elsewhere~\cite{EHT}.

%\section*{acknowledgement}

We thank Wilfried Buchm\"uller for comments. 
M.E. is grateful to Shuntaro Nakamura and 
Masahiro Yamaguchi for private communication. 
M.E. and F.T.  would like to thank the Japan Society for Promotion of 
Science for financial support.

{\it Note Added}: Shortly after our letter, Ref.~\cite{Nakamura:2006uc}
appeared, pointing out the same problem. They also
discussed  the gravitino much heavier than $100$TeV
as a possible solution to the moduli-induced gravitino problem.
More recently, Ref.~\cite{Dine:2006ii} has pointed out $G_X$ can be
much smaller than $R^{-1}$ in some situations.
However, for general non-renormalizable
couplings in $K$, our estimate on the gravitino production still holds.
See \cite{EHT2} for details.

\end{document}